\documentclass{PoS}

\let\oldbibliography\thebibliography
\renewcommand{\thebibliography}[1]{\oldbibliography{#1}
\setlength{\baselineskip}{9.5pt}
\setlength{\itemsep}{0pt}} %Reducing spacing in the bibliography.

\newcommand{\MSb}{{\overline{\rm MS}}}

\title{Light-cone PDFs from Lattice QCD}

\ShortTitle{Light-cone PDFs from Lattice QCD}

\author{Constantia Alexandrou$^{ab}$,
  \speaker{Krzysztof Cichy}$^{c}$,
        Martha Constantinou$^{d}$,  
	Kyriakos Hadjiyiannakou$^{b}$, 
	Karl Jansen$^{e}$,
	Aurora Scapellato$^{a,f}$,
	Fernanda Steffens$^{g}$
\\
\\
        \llap{$^a$}Department of Physics, University of Cyprus, P.O. Box 20537, 1678 Nicosia, Cyprus\\
	\llap{$^b$}Computation-based Science and Technology
	Research Center, Cyprus Institute, 20 Kavafi Str.,
	Nicosia 2121, Cyprus\\
	\llap{$^c$}Faculty of Physics, Adam Mickiewicz
	University, Umultowska 85, 61-614 Pozna\'{n}, Poland\\
	E-mail: \email{kcichy@amu.edu.pl}\\
	\llap{$^d$}Temple University, 1925 N. 12th Street,
	Philadelphia, PA 19122, USA\\
	\llap{$^e$}John von Neumann Institute for Computing (NIC), DESY, Platanenallee 6, D-15738 Zeuthen, Germany\\
	\llap{$^f$}University of Wuppertal, Gau\ss str. 20, 42119 Wuppertal, Germany\\
	\llap{$^g$}Institut f\"{u}r Strahlen- und Kernphysik, Rheinische Friedrich-Wilhelms-Universit\"{a}t Bonn, Nussallee 14-16, 53115 Bonn
        }
        
\abstract{Using the approach proposed a few years ago by X. Ji, it has become feasible to extract parton distribution functions (PDFs) from lattice QCD, a task thought to be extremely difficult before Ji's proposal. In this talk, we discuss this approach, in particular different systematic effects that need to be controlled to ultimately have precise determinations of PDFs. Special attention is paid to the analysis of excited states. We emphasize that it is crucial to control excited states contamination and we show an analysis thereof for our lattice data, used to calculate quasi-PDFs and finally light-cone PDFs in the second part of this proceeding (C.\ Alexandrou et al., \emph{Quasi-PDFs from Twisted mass fermions at the physical point}).}

\FullConference{The 36th Annual International Symposium on Lattice Field Theory - LATTICE2018\\
		22-28 July, 2018\\
		Michigan State University, East Lansing, Michigan, USA.}

\begin{document}

\section{Introduction}
\vspace*{-3mm}
\noindent Collinear parton distribution functions (PDFs) quantify non-perturbative effects in hadronic cross sections and are vital for analysis of scattering experiments such as the ones at the Large Hadron Collider. Even though they result from the QCD Lagrangian, their direct computation is a difficult task. In principle, lattice offers an ideal formulation to study non-perturbative physics. However, PDFs are defined on the light cone, which makes it impossible to translate their definition directly to Euclidean spacetime that lattice QCD requires.
For this reason, lattice computations were for a long time restricted to the low moments of PDFs that can be expressed as matrix elements of local operators.
The interest in getting the full Bjorken-$x$ dependence of PDFs was rather recently revived after the seminal proposal of Ji \cite{Ji:2013dva} to calculate so-called quasi-distributions, expressed in the frame of a boosted nucleon.
Such quasi-distributions can then be matched perturbatively to their light-cone counterparts, utilizing the fact that both kinds of distributions share the same infrared physics and they differ only in the ultraviolet.
The systematic procedure of matching is referred to as large momentum effective theory (LaMET).

Since Ji's paper, there has been a plethora of studies of this approach, both theoretical and numerical.
Renormalizability of quasi-PDFs was established \cite{Ishikawa:2017faj,Ji:2017oey} and both perturbative and non-perturbative renormalization prescriptions were developed \cite{Constantinou:2017sej,Alexandrou:2017huk,Chen:2017mzz,Green:2017xeu,Spanoudes:2018zya}.
Matching between quasi-PDFs renormalized in different schemes and light-cone PDFs was considered \cite{Xiong:2013bka,Chen:2016fxx,Wang:2017qyg,Stewart:2017tvs,Izubuchi:2018srq,Alexandrou:2018pbm} and target mass corrections were calculated \cite{Chen:2016utp}.
Further theoretical aspects were clarified in Refs.\ \cite{Briceno:2017cpo,Briceno:2018lfj,Ji:2017rah,Radyushkin:2018nbf,Karpie:2018zaz}.
The numerical results were presented in Refs.\ \cite{Lin:2014zya,Alexandrou:2015rja,Chen:2016utp,Alexandrou:2016jqi,Alexandrou:2017huk,Chen:2017mzz,Liu:2018uuj,Alexandrou:2018pbm,Alexandrou:2018eet,Chen:2018xof,Lin:2018qky,Liu:2018hxv}, with very encouraging prospects of a fully controlled result in the future. 
Moreover, related direct approaches were also proposed and investigated \cite{Monahan:2016bvm,Radyushkin:2017cyf,Orginos:2017kos,Ma:2017pxb}. 

\vspace*{-5mm}
\section{Quasi-PDFs and light-cone PDFs}
\vspace*{-3mm}
\noindent Quasi-PDFs, $\tilde{q}(x,P)$, are defined as Fourier transforms of the non-local matrix elements $h_\Gamma(z,P)$ $= \langle P\vert \, \overline{\psi}(0,z)\,\Gamma W(z)\,\psi(0,0)\,\vert P\rangle$ for the nucleon boosted to momentum $P=(P_0,0,0,P_3)$, with the type of PDF determined by the Dirac structure $\Gamma$ ($\Gamma{=}\gamma_\mu$ -- unpolarized, $\Gamma{=}\gamma_5\gamma_\mu$ -- helicity and $\Gamma{=}\sigma_{\mu\nu}$ -- transversity) and with $W(z)$ being the Wilson line of length $z$, taken in the spatial direction of the nucleon boost:
\begin{equation}
\label{eq:quasi}
\tilde{q}(x,P){=}\hspace*{-0.1cm}\int_{-\infty}^{+\infty}\hspace*{-0.1cm}\frac{dz}{4\pi}\,
e^{-ixP_3z}\,h_\Gamma(z,P).
\end{equation}
The bare matrix elements $h_\Gamma(z,P)$ are calculated on the lattice as a ratio of suitable three-point and two-point functions, for details about the technical details of the computation, see the next section and the supplemental material of Ref.\ \cite{Alexandrou:2018pbm}. 
$h_\Gamma(z,P)$ are subject to logarithmic and power divergences and hence need renormalization, for details see Ref.\ \cite{Alexandrou:2017huk}, where a non-perturbative renormalization prescription has been developed for the first time.
The Fourier transform of renormalized matrix elements, i.e.\ the renormalized quasi-PDF, can be matched onto light-cone PDFs in the framework of LaMET.
In our work, we choose to adopt a two-step procedure.
First, the matrix elements in the RI$'$ scheme \cite{Alexandrou:2017huk} are perturbatively converted to the $\MSb$ scheme and evolved to a reference scale of 2 GeV \cite{Constantinou:2017sej}.
Second, the matching procedure brings the renormalized quasi-PDF to the light-cone one, without changing the scheme and scale \cite{Alexandrou:2018pbm}.
Finally, target mass corrections are applied using formulae of Ref.\ \cite{Chen:2016utp}.

\vspace*{-5mm}
\section{Lattice setup}
\vspace*{-3mm}
\noindent We use a gauge field configurations ensemble with two degenerate clover-improved Wilson twisted mass quarks of physical masses and Iwasaki gluons \cite{Abdel-Rehim:2015pwa}.
The lattice size is $48^3\times 96$ and the lattice spacing $a{=}0.0938(3)(2)$~fm, which gives a spatial extent of $L=4.5$~fm.

The nucleon boosts used in our work are $\frac{6\pi}{L}$, $\frac{8\pi}{L}$ and $\frac{10\pi}{L}$, i.e.\ $0.83,\,1.11,\,1.38$~GeV, respectively.
The largest momentum that can be reached with realistic computer resources is limited by the separation between source and sink, $t_s$, in the three-point function.
We show below that $t_s$ above 1 fm in physical units is needed to suppress enough the contribution of excited states, if aiming at a precision of 10\%.
In our study, we consider source-sink separations of 0.75, 0.84, 0.93 and 1.12 fm and establish excited states suppression at the desired level by comparing single-state, two-state and summation fits.
Our statistics at the largest $t_s$ corresponds to 4800-9600 measurements (depending on the Dirac structure) at $P_3=\frac{6\pi}{L}$, 38250 measurements at $P_3=\frac{8\pi}{L}$ and 72990 measurements at $P_3=\frac{10\pi}{L}$. The increase of statistics with increasing momentum results from the desire to obtain similar precision of bare matrix elements at all momenta under the decaying signal-to-noise ratio at larger $P_3$.
The decay of the signal is still exponential, despite using the technique of momentum smearing \cite{Bali:2016lva} with optimized parameters.
In addition to this method, we use other techniques to optimize the signal, such as low precision inversions with correction of the resulting bias.

\vspace*{-5mm}
\section{Systematic effects -- general discussion}
\vspace*{-3mm}
\noindent A reliable computation of PDFs within the framework of LaMET needs, apart from small statistical uncertainties, control over numerous sources of systematic effects.
Here, we discuss the status of our calculation from this point of view.

One of the biggest challenges is to reach nucleon boosts large enough to make contact to the light-cone frame, equivalently the infinite momentum frame, with enough suppression of excited states contamination at the desired precision level.
Excited states can never be eliminated altogether -- they can only become subleading with respect to other sources of uncertainty.
This aspect is essential for having final reliable PDFs and we devote a separate section to it, see below.

Other systematic effects can be divided into typical lattice systematics and ones related specifically to the considered observable.
For the former, we mention the role of the pion mass, cut-off effects, finite volume effects (FVE) and lattice artifacts in the renormalization procedure.
Among these, the pion mass was the first that we fully addressed by performing computations on an ensemble with light quark masses tuned to reproduce the physical pion mass.
The second one, that we are in the process of analyzing, is the role of lattice artifacts in Z-factors.
Our renormalization programme, introduced in Ref.\ \cite{Alexandrou:2017huk}, relies on the non-perturbative RI$'$ scheme and Z-factors computed at different renormalization scales can be differently contaminated by the lattice breaking of rotational symmetry.
The discretization effects induced by this can be computed in lattice perturbation theory and subtracted, making the final Z-factors estimates more reliable.
It has been demonstrated that the procedure works very efficiently for local Z-factors \cite{Alexandrou:2015sea}.
For cut-off effects and FVE in bare matrix elements, an explicit computation is required using ensembles with different lattice spacings and volumes.
Nevertheless, there are indirect premises that both of these effects should not be large -- they were typically found to be small in hadron structure computations and moreover, we find that the continuum dispersion relation is satisfied on the lattice up to the largest employed momentum \cite{Alexandrou:2018pbm}.

The quasi-PDF specific systematic effects include, most notably, higher-twist effects from using finite nucleon boost.
These can be suppressed by using large enough momentum and we notice signs of convergence when increasing boost from 0.83 GeV to 1.38 GeV \cite{Alexandrou:2018pbm}.
Nevertheless, we observe unphysical behavior in the data in some regions of $x$, which signifies the nucleon momentum is still too small.
As a solution to this problem, the derivative method was proposed \cite{Lin:2017ani}, which consists in rewriting the Fourier transform using integration by parts and ignoring the surface term.
However, it is clear that this does not solve the problem, but only hides its presence.
Moreover, ignoring the surface term yields uncontrolled behavior for $|x|\lesssim0.2$, both in the quark and antiquark part, and introduces additional discretization effects from the discretization of the derivative of matrix elements.
A more promising way of alleviating effects of working at finite momentum may be to compute the higher-twist effects explicitly, see an analysis of these in quasi-PDFs in Ref.\ \cite{Braun:2018brg}.
Another systematic effect that has to be addressed for a reliable extraction of PDFs is related to truncating perturbative expressions for matching of quasi-PDFs to light-cone PDFs.
The conversion between the intermediate lattice renormalization scheme, like a variant of RI-MOM, and the $\MSb$ scheme, desired in phenomenological applications, necessarily proceeds perturbatively, as well as the evolution to the reference scale, like 2 GeV.
All of these perturbative expressions are presently truncated at one-loop, without reliable estimates of higher-loop effects.
The magnitude of one-loop effects is rather large and hence, a two-loop computation is mandatory and can lead to sizable contributions.

\vspace*{-5mm}
\section{Excited states effects}
\vspace*{-3mm}
\noindent One of the crucial effects in lattice computation of hadron structure observables is the excited states contamination.
The lattice two-point and three-point correlators can be decomposed into a series of terms containing exponentials of masses of states excited from the vacuum, with appropriate quantum numbers, multiplied by Euclidean times expressing the temporal separation between the source and the sink or between the source/sink and the operator insertion time.
For reliable statements about the ground state hadron, all these separations have to be large enough to suppress exponentials corresponding to excited states.
The history of lattice QCD hadron structure calculations has shown that excited states contamination can prove to be a major effect, hindering the safe extraction of even simple quantities, like $\langle x\rangle_{u-d}$ or $g_A$, see e.g.\ \cite{Abdel-Rehim:2015owa,Gupta:2018qil}.
Moreover, boosting the nucleon increases the role of excited states and a careful analysis is needed to reliably prove they are subdominating with respect to other sources of systematic uncertainties.

\begin{figure}[h!]
\begin{center}
   \includegraphics[width=0.47\textwidth]{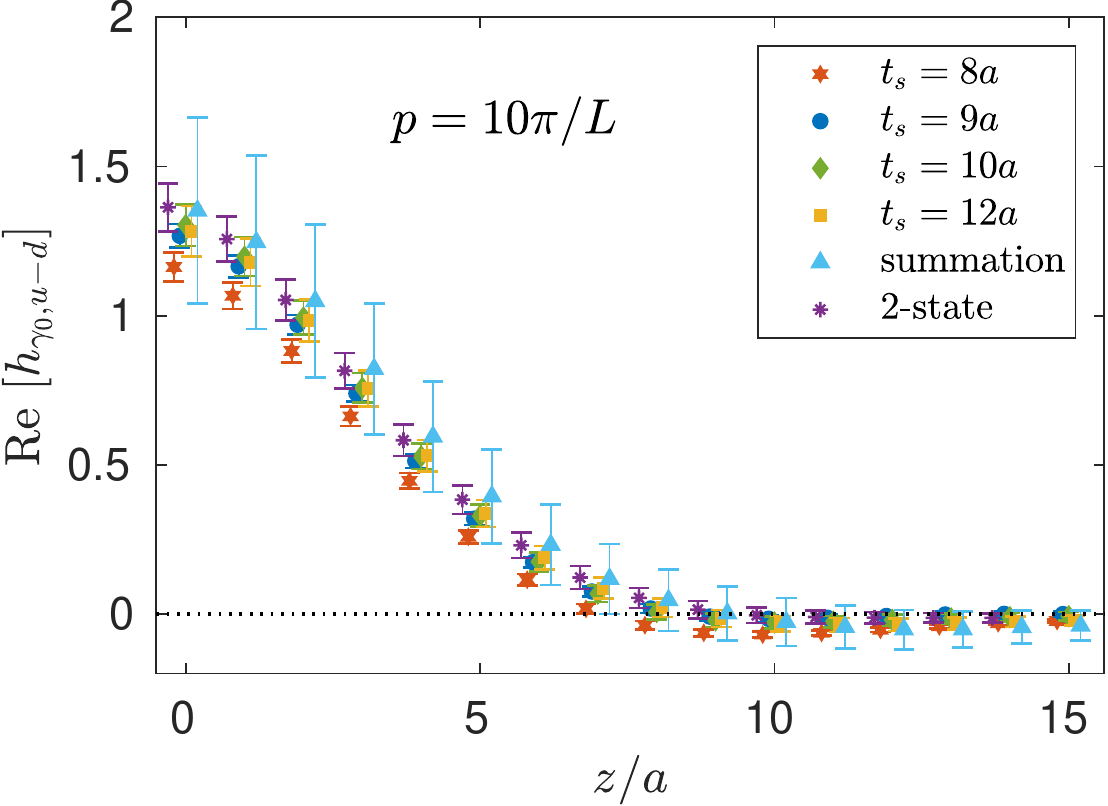}
   \includegraphics[width=0.47\textwidth]{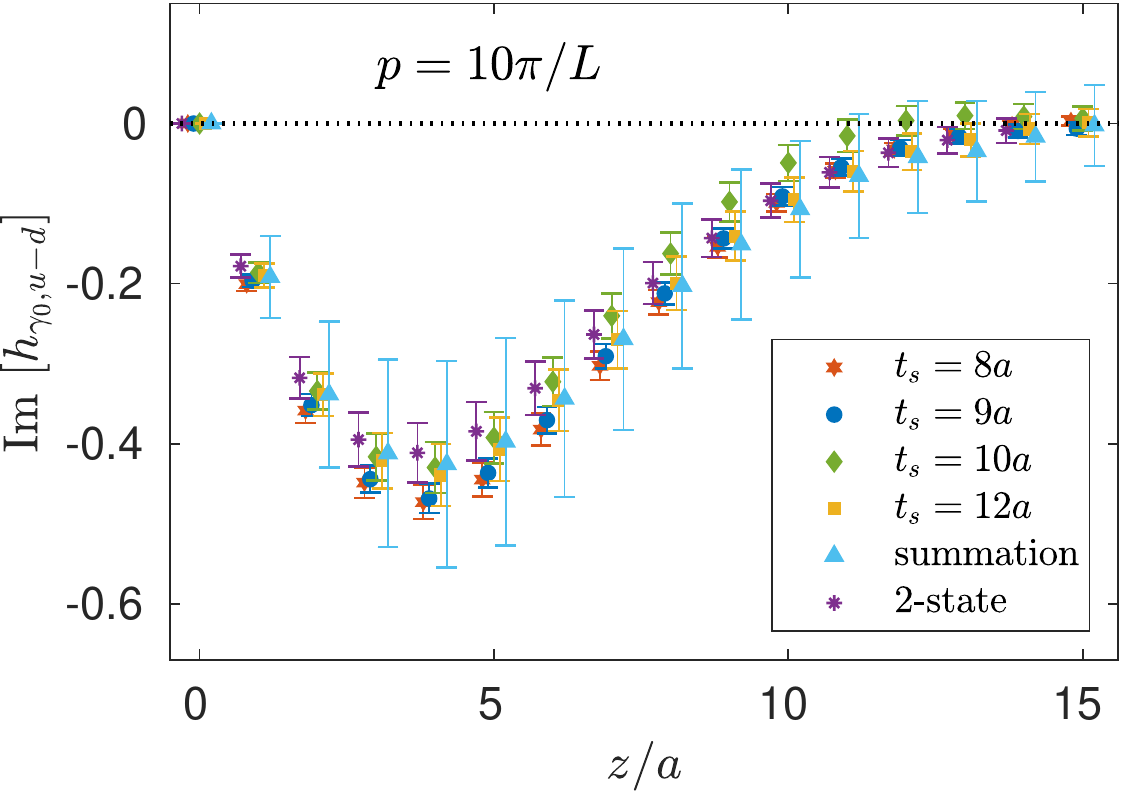}
   \includegraphics[width=0.47\textwidth]{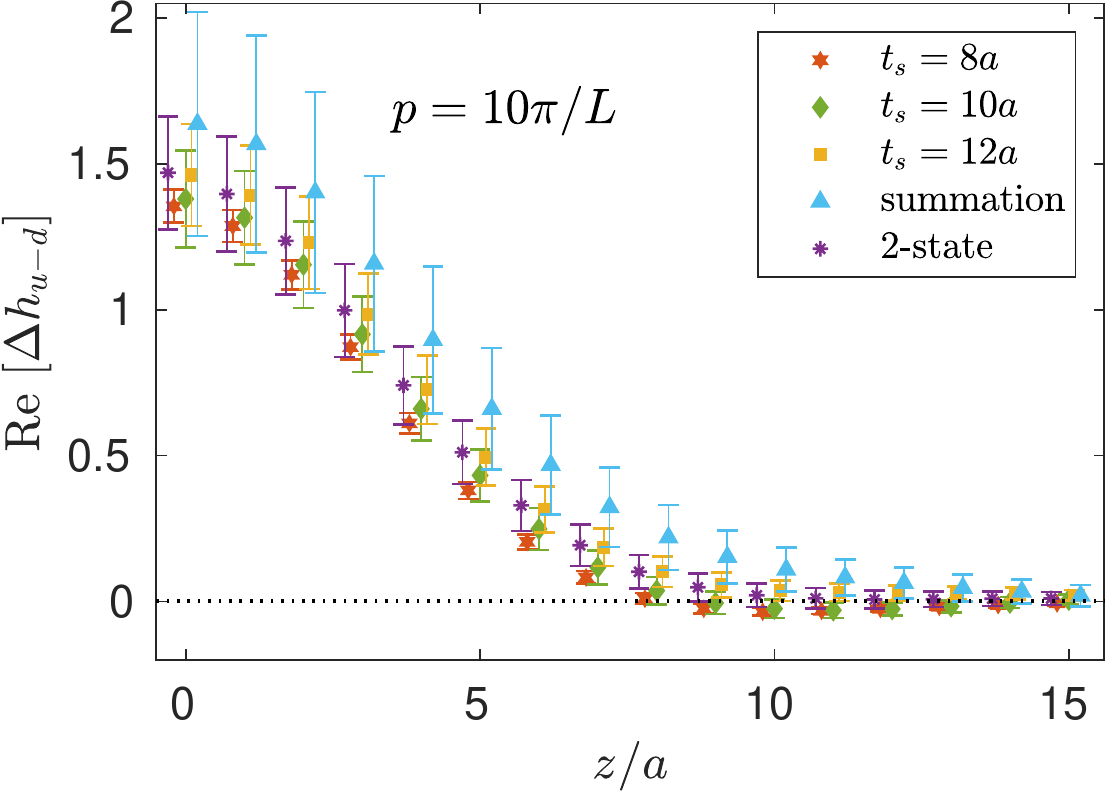}
   \includegraphics[width=0.47\textwidth]{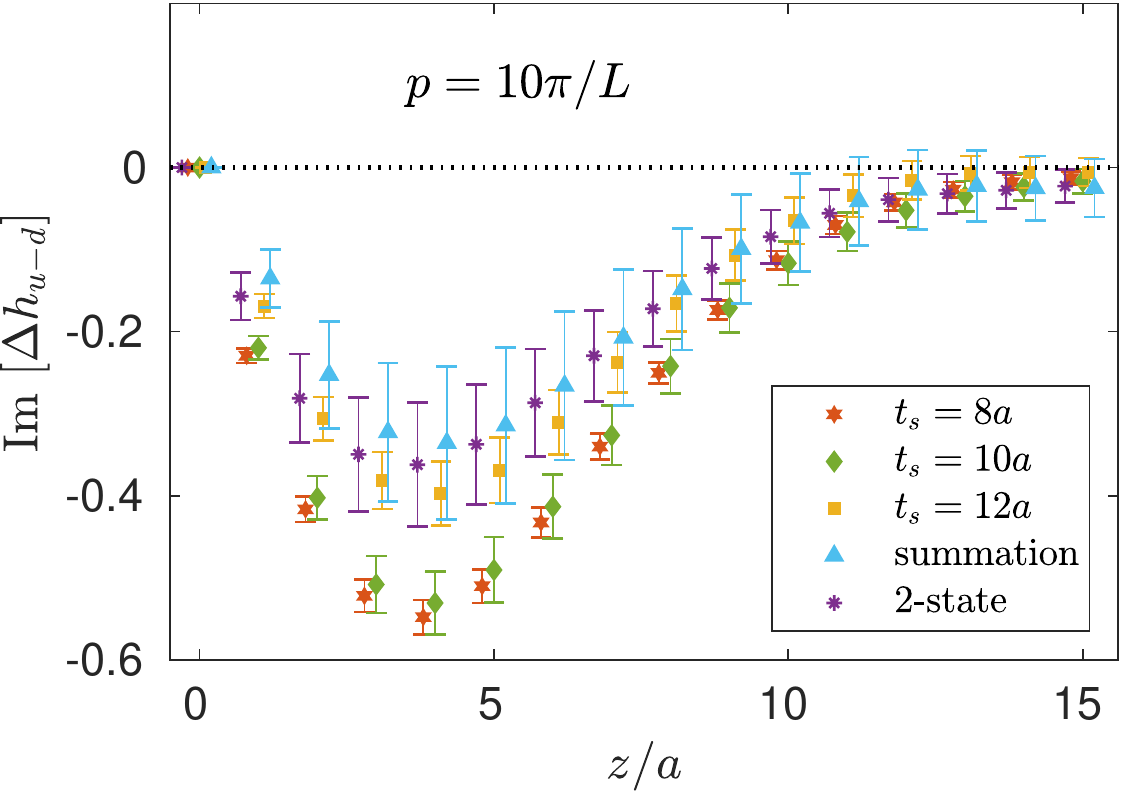}
   \includegraphics[width=0.47\textwidth]{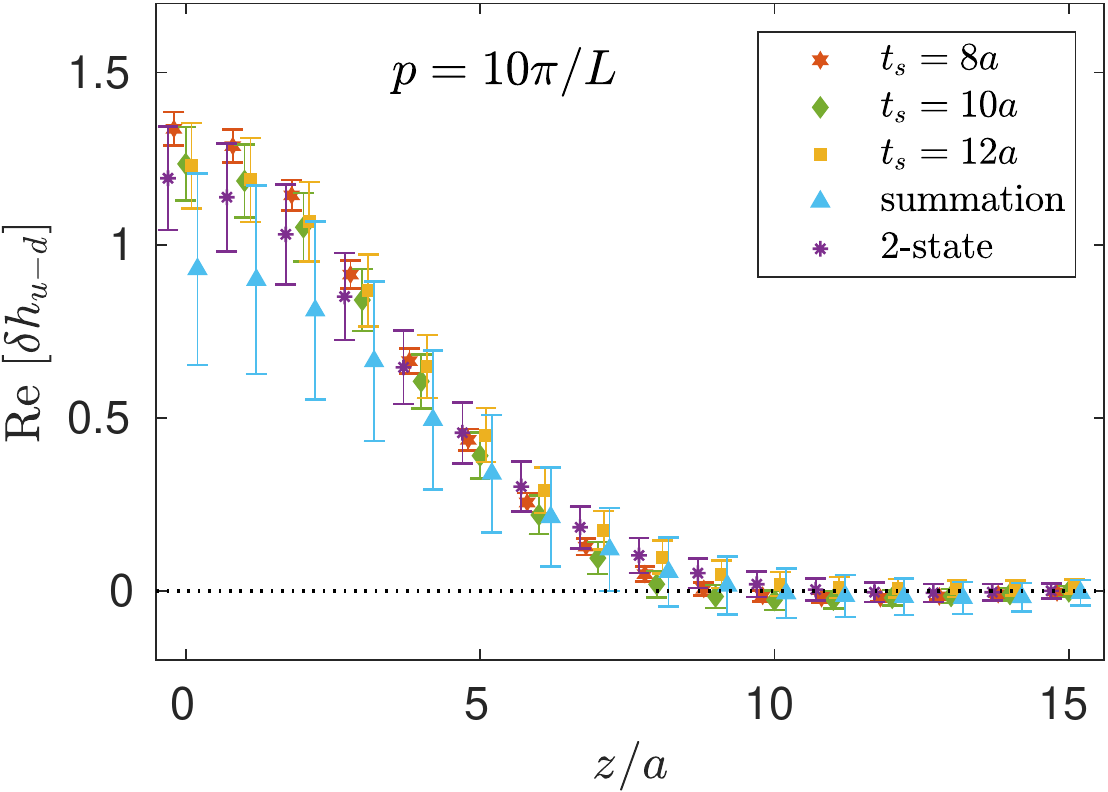}
   \includegraphics[width=0.47\textwidth]{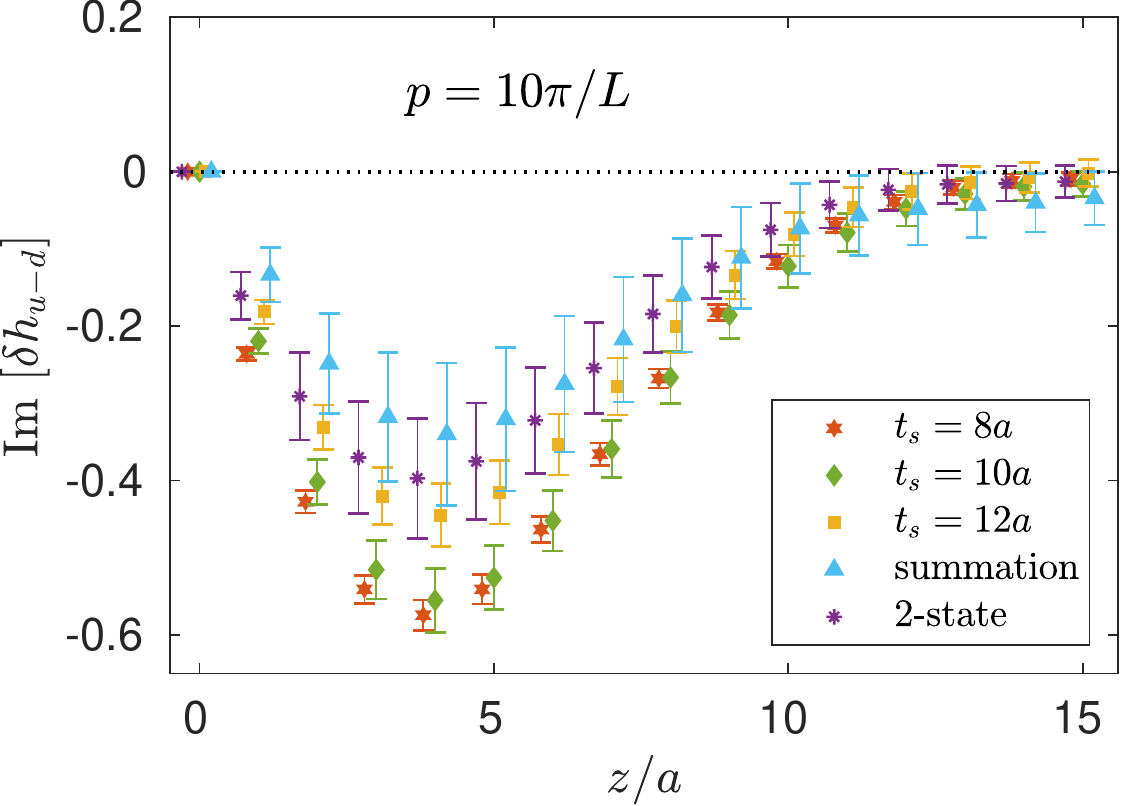}
\caption{Real (left) and imaginary (right) part of the matrix element for different PDFs (top -- unpolarized, middle -- helicity, bottom -- transversity) from the plateau method (points labeled with appropriate $t_s$), the summation method and two-state fits (to all $t_s$). Nucleon momentum is $10\pi/L\simeq 1.38$ GeV.}
\label{fig:excited}
\end{center}
\end{figure}

In this work, we analyze the role of excited states comparing three kinds of methods of extracting bare matrix elements: single-state (plateau) fits, two-state fits and the summation method. Details of the study will be given in a forthcoming publication.
We consider 4 source-sink separations for the unpolarized case ($t_s=8a,9a,10a,12a$; $t_s\approx0.75,0.84,0.93,1.12$ fm in physical units) and we increase statistics to keep the statistical uncertainties approximately equal for all separations, i.e.\ we use 4320, 8820, 9000, 72990 measurements, respectively.
For both polarized cases, we use 3 values of $t_s=8a,10a,12a$, with 3240, 7920 and 72990 measurements.

The summation method has much larger statistical errors than the other ones and is inconclusive for the unpolarized matrix elements.
However, in the polarized ones, the real part of the matrix elements for large $z/a$ and the imaginary part for small $z/a$ is compatible only with two-state fits and the plateau method at $t_s=12a$. 
The two-state fits always yield compatible results with $t_s=12a$ plateau fits, while there is clear tension with smaller source-sink separations for all Dirac structures in some regions of $z/a$ in the imaginary part, especially for helicity and transversity.
Given our statistical precision, we have, thus, established ground state dominance at the approx.\ 10\% level for source-sink separation 1.12 fm.
This is in accordance with earlier hadron structure studies.
However, aiming at a higher precision, the conclusion might not hold, i.e.\ increase of $t_s$ may be needed to sufficiently control excited states.
The same is true when attempting simulations with larger momenta and we emphasize that establishing compatibility between the three methods of extracting bare matrix elements is a prerequisite for a reliable analysis.

Having determined that $t_s=12a$ is safe from excited states at the 10\% level, we can proceed to further steps required to obtain light-cone PDFs, i.e.\ renormalization and matching, described in the second part of these proceedings \cite{ProcAurora2018}.

\vspace*{-4mm}
\section*{Acknowledgments}
\vspace*{-3mm}
\noindent This work has received funding from the European Union's Horizon 2020 research and innovation
programme under the Marie Sk\l{}odowska-Curie grant agreement no.\ 642069 (HPC-LEAP). K.C.\
is supported by the National Science Centre grant SONATA BIS no.\ 2016/22/E/ST2/00013. F.S.\
is funded by the Deutsche Forschungsgemeinschaft (DFG) project no.\ 392578569. 
M.C.\ acknowledges financial support by the U.S.\ Department of Energy, Office of Nuclear Physics, within
the framework of the TMD Topical Collaboration and by the National Science Foundation under Grant no.\ PHY-1714407. 
This research used resources of the Oak Ridge Leadership Computing Facility, which is a DOE Office of Science User Facility supported under contract DE-AC05-00OR22725, J\"ulich Supercomputing Centre, Prometheus supercomputer at the Academic Computing Centre Cyfronet AGH in Cracow, Okeanos supercomputer at the Interdisciplinary Centre for Mathematical and Computational Modelling in Warsaw, Eagle supercomputer at the Poznan Supercomputing and Networking Center.

\vspace*{-3mm}
\bibliographystyle{JHEP}
\bibliography{references}

\providecommand{\href}[2]{#2}\begingroup\raggedright\begin{thebibliography}{10}

\bibitem{Ji:2013dva}
X.~Ji, {\it {Parton Physics on a Euclidean Lattice}},  {\em Phys.Rev.Lett.}
  {\bf 110} (2013) 262002, [\href{http://arxiv.org/abs/1306.1539}{{\tt
  arXiv:1306.1539}}].

\bibitem{Ishikawa:2017faj}
T.~Ishikawa, Y.-Q. Ma, J.-W. Qiu, and S.~Yoshida, {\it {Renormalizability of
  quasiparton distribution functions}},  {\em Phys. Rev.} {\bf D96} (2017),
  no.~9 094019, [\href{http://arxiv.org/abs/1707.03107}{{\tt
  arXiv:1707.03107}}].

\bibitem{Ji:2017oey}
X.~Ji, J.-H. Zhang, and Y.~Zhao, {\it {Renormalization in Large Momentum
  Effective Theory of Parton Physics}},  {\em Phys. Rev. Lett.} {\bf 120}
  (2018), no.~11 112001, [\href{http://arxiv.org/abs/1706.08962}{{\tt
  arXiv:1706.08962}}].

\bibitem{Constantinou:2017sej}
M.~Constantinou and H.~Panagopoulos, {\it {Perturbative renormalization of
  quasi-parton distribution functions}},  {\em Phys. Rev.} {\bf D96} (2017),
  no.~5 054506, [\href{http://arxiv.org/abs/1705.11193}{{\tt
  arXiv:1705.11193}}].

\bibitem{Alexandrou:2017huk}
C.~Alexandrou, K.~Cichy, M.~Constantinou, K.~Hadjiyiannakou, K.~Jansen,
  H.~Panagopoulos, and F.~Steffens, {\it {A complete non-perturbative
  renormalization prescription for quasi-PDFs}},  {\em Nucl. Phys.} {\bf B923}
  (2017) 394--415, [\href{http://arxiv.org/abs/1706.00265}{{\tt
  arXiv:1706.00265}}].

\bibitem{Chen:2017mzz}
J.-W. Chen, T.~Ishikawa, L.~Jin, H.-W. Lin, Y.-B. Yang, J.-H. Zhang, and
  Y.~Zhao, {\it {Parton distribution function with nonperturbative
  renormalization from lattice QCD}},  {\em Phys. Rev.} {\bf D97} (2018), no.~1
  014505, [\href{http://arxiv.org/abs/1706.01295}{{\tt arXiv:1706.01295}}].

\bibitem{Green:2017xeu}
J.~Green, K.~Jansen, and F.~Steffens, {\it {Nonperturbative Renormalization of
  Nonlocal Quark Bilinears for Parton Quasidistribution Functions on the
  Lattice Using an Auxiliary Field}},  {\em Phys. Rev. Lett.} {\bf 121} (2018),
  no.~2 022004, [\href{http://arxiv.org/abs/1707.07152}{{\tt
  arXiv:1707.07152}}].

\bibitem{Spanoudes:2018zya}
G.~Spanoudes and H.~Panagopoulos, {\it {Renormalization of Wilson-line
  operators in the presence of nonzero quark masses}},  {\em Phys. Rev.} {\bf
  D98} (2018), no.~1 014509, [\href{http://arxiv.org/abs/1805.01164}{{\tt
  arXiv:1805.01164}}].

\bibitem{Xiong:2013bka}
X.~Xiong, X.~Ji, J.-H. Zhang, and Y.~Zhao, {\it {One-loop matching for parton
  distributions: Nonsinglet case}},  {\em Phys.Rev.} {\bf D90} (2014), no.~1
  014051, [\href{http://arxiv.org/abs/1310.7471}{{\tt arXiv:1310.7471}}].

\bibitem{Chen:2016fxx}
J.-W. Chen, X.~Ji, and J.-H. Zhang, {\it {Improved quasi parton distribution
  through Wilson line renormalization}},  {\em Nucl. Phys.} {\bf B915} (2017)
  1--9, [\href{http://arxiv.org/abs/1609.08102}{{\tt arXiv:1609.08102}}].

\bibitem{Wang:2017qyg}
W.~Wang, S.~Zhao, and R.~Zhu, {\it {Gluon quasidistribution function at one
  loop}},  {\em Eur. Phys. J.} {\bf C78} (2018), no.~2 147,
  [\href{http://arxiv.org/abs/1708.02458}{{\tt arXiv:1708.02458}}].

\bibitem{Stewart:2017tvs}
I.~W. Stewart and Y.~Zhao, {\it {Matching the quasiparton distribution in a
  momentum subtraction scheme}},  {\em Phys. Rev.} {\bf D97} (2018), no.~5
  054512, [\href{http://arxiv.org/abs/1709.04933}{{\tt arXiv:1709.04933}}].

\bibitem{Izubuchi:2018srq}
T.~Izubuchi, X.~Ji, L.~Jin, I.~W. Stewart, and Y.~Zhao, {\it {Factorization
  Theorem Relating Euclidean and Light-Cone Parton Distributions}},  {\em Phys.
  Rev.} {\bf D98} (2018), no.~5 056004,
  [\href{http://arxiv.org/abs/1801.03917}{{\tt arXiv:1801.03917}}].

\bibitem{Alexandrou:2018pbm}
C.~Alexandrou, K.~Cichy, M.~Constantinou, K.~Jansen, A.~Scapellato, and
  F.~Steffens, {\it {Light-Cone Parton Distribution Functions from Lattice
  QCD}},  {\em Phys. Rev. Lett.} {\bf 121} (2018), no.~11 112001,
  [\href{http://arxiv.org/abs/1803.02685}{{\tt arXiv:1803.02685}}].

\bibitem{Chen:2016utp}
J.-W. Chen, S.~D. Cohen, X.~Ji, H.-W. Lin, and J.-H. Zhang, {\it {Nucleon
  Helicity and Transversity Parton Distributions from Lattice QCD}},  {\em
  Nucl. Phys.} {\bf B911} (2016) 246--273,
  [\href{http://arxiv.org/abs/1603.06664}{{\tt arXiv:1603.06664}}].

\bibitem{Briceno:2017cpo}
R.~A. Brice{\~n}o, M.~T. Hansen, and C.~J. Monahan, {\it {Role of the Euclidean
  signature in lattice calculations of quasidistributions and other nonlocal
  matrix elements}},  {\em Phys. Rev.} {\bf D96} (2017), no.~1 014502,
  [\href{http://arxiv.org/abs/1703.06072}{{\tt arXiv:1703.06072}}].

\bibitem{Briceno:2018lfj}
R.~A. Brice{\~n}o, J.~V. Guerrero, M.~T. Hansen, and C.~J. Monahan, {\it
  {Finite-volume effects due to spatially nonlocal operators}},  {\em Phys.
  Rev.} {\bf D98} (2018), no.~1 014511,
  [\href{http://arxiv.org/abs/1805.01034}{{\tt arXiv:1805.01034}}].

\bibitem{Ji:2017rah}
X.~Ji, J.-H. Zhang, and Y.~Zhao, {\it {More On Large-Momentum Effective Theory
  Approach to Parton Physics}},  {\em Nucl. Phys.} {\bf B924} (2017) 366--376,
  [\href{http://arxiv.org/abs/1706.07416}{{\tt arXiv:1706.07416}}].

\bibitem{Radyushkin:2018nbf}
A.~V. Radyushkin, {\it {Structure of parton quasi-distributions and their
  moments}},  \href{http://arxiv.org/abs/1807.07509}{{\tt arXiv:1807.07509}}.

\bibitem{Karpie:2018zaz}
J.~Karpie, K.~Orginos, and S.~Zafeiropoulos, {\it {Moments of Ioffe time parton
  distribution functions from non-local matrix elements}},
  \href{http://arxiv.org/abs/1807.10933}{{\tt arXiv:1807.10933}}.

\bibitem{Lin:2014zya}
H.-W. Lin, J.-W. Chen, S.~D. Cohen, and X.~Ji, {\it {Flavor Structure of the
  Nucleon Sea from Lattice QCD}},  {\em Phys. Rev.} {\bf D91} (2015) 054510,
  [\href{http://arxiv.org/abs/1402.1462}{{\tt arXiv:1402.1462}}].

\bibitem{Alexandrou:2015rja}
C.~Alexandrou, K.~Cichy, V.~Drach, E.~Garcia-Ramos, K.~Hadjiyiannakou,
  K.~Jansen, F.~Steffens, and C.~Wiese, {\it {Lattice calculation of parton
  distributions}},  {\em Phys. Rev.} {\bf D92} (2015) 014502,
  [\href{http://arxiv.org/abs/1504.07455}{{\tt arXiv:1504.07455}}].

\bibitem{Alexandrou:2016jqi}
C.~Alexandrou, K.~Cichy, M.~Constantinou, K.~Hadjiyiannakou, K.~Jansen,
  F.~Steffens, and C.~Wiese, {\it {Updated Lattice Results for Parton
  Distributions}},  {\em Phys. Rev.} {\bf D96} (2017), no.~1 014513,
  [\href{http://arxiv.org/abs/1610.03689}{{\tt arXiv:1610.03689}}].

\bibitem{Liu:2018uuj}
Y.-S. Liu, J.-W. Chen, L.~Jin, H.-W. Lin, Y.-B. Yang, J.-H. Zhang, and Y.~Zhao,
  {\it {Unpolarized quark distribution from lattice QCD: A systematic analysis
  of renormalization and matching}},
  \href{http://arxiv.org/abs/1807.06566}{{\tt arXiv:1807.06566}}.

\bibitem{Alexandrou:2018eet}
C.~Alexandrou, K.~Cichy, M.~Constantinou, K.~Jansen, A.~Scapellato, and
  F.~Steffens, {\it {Transversity parton distribution functions from lattice
  QCD}},  \href{http://arxiv.org/abs/1807.00232}{{\tt arXiv:1807.00232}}.

\bibitem{Chen:2018xof}
J.-W. Chen, L.~Jin, H.-W. Lin, Y.-S. Liu, Y.-B. Yang, J.-H. Zhang, and Y.~Zhao,
  {\it {Lattice Calculation of Parton Distribution Function from LaMET at
  Physical Pion Mass with Large Nucleon Momentum}},
  \href{http://arxiv.org/abs/1803.04393}{{\tt arXiv:1803.04393}}.

\bibitem{Lin:2018qky}
H.-W. Lin, J.-W. Chen, L.~Jin, Y.-S. Liu, Y.-B. Yang, J.-H. Zhang, and Y.~Zhao,
  {\it {Spin-Dependent Parton Distribution Function with Large Momentum at
  Physical Pion Mass}},  \href{http://arxiv.org/abs/1807.07431}{{\tt
  arXiv:1807.07431}}.

\bibitem{Liu:2018hxv}
Y.-S. Liu, J.-W. Chen, L.~Jin, R.~Li, H.-W. Lin, Y.-B. Yang, J.-H. Zhang, and
  Y.~Zhao, {\it {Nucleon Transversity Distribution at the Physical Pion Mass
  from Lattice QCD}},  \href{http://arxiv.org/abs/1810.05043}{{\tt
  arXiv:1810.05043}}.

\bibitem{Monahan:2016bvm}
C.~Monahan and K.~Orginos, {\it {Quasi parton distributions and the gradient
  flow}},  {\em JHEP} {\bf 03} (2017) 116,
  [\href{http://arxiv.org/abs/1612.01584}{{\tt arXiv:1612.01584}}].

\bibitem{Radyushkin:2017cyf}
A.~V. Radyushkin, {\it {Quasi-parton distribution functions, momentum
  distributions, and pseudo-parton distribution functions}},  {\em Phys. Rev.}
  {\bf D96} (2017), no.~3 034025, [\href{http://arxiv.org/abs/1705.01488}{{\tt
  arXiv:1705.01488}}].

\bibitem{Orginos:2017kos}
K.~Orginos, A.~Radyushkin, J.~Karpie, and S.~Zafeiropoulos, {\it {Lattice QCD
  exploration of parton pseudo-distribution functions}},  {\em Phys. Rev.} {\bf
  D96} (2017), no.~9 094503, [\href{http://arxiv.org/abs/1706.05373}{{\tt
  arXiv:1706.05373}}].

\bibitem{Ma:2017pxb}
Y.-Q. Ma and J.-W. Qiu, {\it {Exploring Partonic Structure of Hadrons Using ab
  initio Lattice QCD Calculations}},  {\em Phys. Rev. Lett.} {\bf 120} (2018),
  no.~2 022003, [\href{http://arxiv.org/abs/1709.03018}{{\tt
  arXiv:1709.03018}}].

\bibitem{Abdel-Rehim:2015pwa}
{\bf ETM} Collaboration, A.~Abdel-Rehim et~al., {\it {First physics results at
  the physical pion mass from $N_f=2$ Wilson twisted mass fermions at maximal
  twist}},  {\em Phys. Rev.} {\bf D95} (2017), no.~9 094515,
  [\href{http://arxiv.org/abs/1507.05068}{{\tt arXiv:1507.05068}}].

\bibitem{Bali:2016lva}
G.~S. Bali, B.~Lang, B.~U. Musch, and A.~Sch{\"a}fer, {\it {Novel quark
  smearing for hadrons with high momenta in lattice QCD}},  {\em Phys. Rev.}
  {\bf D93} (2016), no.~9 094515, [\href{http://arxiv.org/abs/1602.05525}{{\tt
  arXiv:1602.05525}}].

\bibitem{Alexandrou:2015sea}
{\bf ETM} Collaboration, C.~Alexandrou, M.~Constantinou, and H.~Panagopoulos,
  {\it {Renormalization functions for Nf=2 and Nf=4 twisted mass fermions}},
  {\em Phys. Rev.} {\bf D95} (2017), no.~3 034505,
  [\href{http://arxiv.org/abs/1509.00213}{{\tt arXiv:1509.00213}}].

\bibitem{Lin:2017ani}
{\bf LP3} Collaboration, H.-W. Lin, J.-W. Chen, T.~Ishikawa, and J.-H. Zhang,
  {\it {Improved parton distribution functions at the physical pion mass}},
  {\em Phys. Rev.} {\bf D98} (2018), no.~5 054504,
  [\href{http://arxiv.org/abs/1708.05301}{{\tt arXiv:1708.05301}}].

\bibitem{Braun:2018brg}
V.~M. Braun, A.~Vladimirov, and J.-H. Zhang, {\it {Power corrections and
  renormalons in parton quasi-distributions}},
  \href{http://arxiv.org/abs/1810.00048}{{\tt arXiv:1810.00048}}.

\bibitem{Abdel-Rehim:2015owa}
A.~Abdel-Rehim et~al., {\it {Nucleon and pion structure with lattice QCD
  simulations at physical value of the pion mass}},  {\em Phys. Rev.} {\bf D92}
  (2015), no.~11 114513, [\href{http://arxiv.org/abs/1507.04936}{{\tt
  arXiv:1507.04936}}]. [Erratum: Phys. Rev.D93,no.3,039904(2016)].

\bibitem{Gupta:2018qil}
R.~Gupta, Y.-C. Jang, B.~Yoon, H.-W. Lin, V.~Cirigliano, and T.~Bhattacharya,
  {\it {Isovector Charges of the Nucleon from 2+1+1-flavor Lattice QCD}},
  \href{http://arxiv.org/abs/1806.09006}{{\tt arXiv:1806.09006}}.

\bibitem{ProcAurora2018}
C.~Alexandrou, K.~Cichy, M.~Constantinou, K.~Hadjiyiannakou, K.~Jansen,
  A.~Scapellato, and F.~Steffens, {\it {Quasi-PDFs from Twisted mass fermions
  at the physical point}},  in {\em {\pos{PoS(LATTICE2018)095}}}.

\end{thebibliography}\endgroup
%\begin{thebibliography}{99}
%\bibitem{...}
%....

%\end{thebibliography}

\end{document}